# SINGLE-HIT CRITERION IN DAMA/LIBRA DM SEARCH AND DAEMONS – THEY ARE ANYTHING BUT WEAKLY INTERACTING

E.M.Drobyshevski

*Ioffe Physical-Technical Institute, Russian Academy of Sciences, 194021 St-Petersburg, Russia*
*emdrob@mail.ioffe.ru*

Our prediction that the more massive DAMA/LIBRA detector would detect a smaller number of events per unit of mass and time than the DAMA/NaI system has got confirmation. This is easy to understand, because DM objects are by far not the WIMPs of the Galactic halo that interact only weakly with matter but are apparently instead electrically charged Planckian objects, i.e., daemons which fall from Earth-crossing orbits with $V$ = 30-50 km/s and undergo multiple interaction with condensed matter already in its outer layers, on a path of a few tens of cm. Therefore, one should use not compact massive detectors but rather systems with a large surface area, as we did to detect daemons with thin ZnS(Ag) scintillators. There are grounds to believe that correct use of the single-hit criterion in LIBRA should reveal DM particles with $V$ = 30-50 km/s, and subsequently, with $V$ = 10-15 km/s as well.

## 1. Introduction. The DAMA/NaI experiment: its interpretation within the daemon paradigm

A more than 15-year-long experimental search for dark mass (DM) objects of our Galaxy halo has not thus far yielded universally recognized results. It is presently fashionable to search for WIMPs, hypothetical weakly interacting particles with mass $M \sim 10^2$ GeV, predicted by various theories beyond the Standard Model. The accuracy of measurements has increased during this time by nearly 4-5 orders of magnitude compared with the figures originally believed good enough to ensure their detection.

The only result that inspires hope is the annual modulation, with a maximum somewhere in early June, in the number of scintillations with an amplitude $A \approx 0.02$ cpd/kg/keV within a narrow energy interval of 2-6 keV (electronic) equivalent, which was revealed at a confidence level of 6.3σ in the DAMA/NaI experiment at Gran Sasso in 1996-2002. The events are believed to be caused by the Galaxy halo WIMPs while the modulation is assigned to the Earth's motion in an orbit inclined to the direction of the Sun's motion around the Galactic center. The DAMA/NaI detector consisted essentially of nine (3×3) 9.7-kg NaI(Tl) crystals of a high radiation purity, each measures 10.2×10.2×25.4 cm$^3$. They are arranged horizontally and parallel to one another. The energy of $E_0$ = 2 keV corresponds to the lower sensitivity threshold of the system, and scintillations with energies >6 keV do not reveal any modulation in the appearance frequency [1,2].

Attempts of other researchers at reproducing the results of DAMA/NaI with other detectors (LXe, Si, Ge and others) have not thus far been crowned with success (see refs. in [1,2]).

On our side, we undertook a search for daemons, DArk Electric Matter Objects, presumably relic elementary Planckian black holes ($M \sim 10^{19}$ GeV, $r_g \sim 10^{-33}$ cm) carrying a negative electric charge of $\approx 10e$, whose repulsion is counteracted by self-gravitation of the object. Because of the giant mass of each daemon, their flux from the Galactic halo or disk is fairly small, about $f \sim 10^{-13}$ cm$^{-2}$s$^{-1}$ only. This is why we were looking for daemons captured by combined action of the Sun and the Earth from the Galactic disk into Earth-crossing orbits. These daemons build up in Strongly Elongated Earth-Crossing Heliocentric Orbits (SEECHOs), to produce an Earth-hitting flux which, by our estimates [3], is as high as $f \sim 3 \times 10^{-7}$ cm$^{-2}$s$^{-1}$ for a velocity $V \sim 30$-$50$ km/s. After multiple encounters with the Earth, part of these daemons becomes captured into Near-Earth Almost Circular Heliocentric Orbits (NEACHOs), where they build up and from where they fall onto the Earth with $V = 10(11.2)$-$15$ km/s.

As shown by celestial mechanics calculations, assuming an effective cross section of daemon interaction with solar matter $\sigma \approx 10^{-19}$ cm$^2$, the SEECHO objects form a kind of a shadow crossing the Earth's orbit in April-August in the antapex zone, i.e., in the region opposite to the direction of the Sun's motion with respect to the nearest stellar population or interstellar gas clouds. As for the NEACHO daemons, they cross the Earth's orbit primarily close to the equinox regions, actually ~10 days earlier [4].

Our purposeful measurements, both ground-level and underground, performed with thin ZnS(Ag) scintillators, have revealed at a confidence level of $\approx 99.99\%$ the existence of a flux of daemons $f > 10^{-7}$ cm$^{-2}$s$^{-1}$, which fall with $V = 10$-$15$ km/s (actually, we observed for a base of $\approx 30$ cm a shift in time $\Delta t$ of signals within a $\Delta t = 20$-$40$ µs bin in the signal distribution $N(\Delta t)$). The flux was found to vary with $P = 0.5$ y with maxima in March and September [5,6].

As shown by an analysis of possible properties of daemons which were revealed in the course of the experiment (indeed, the time of the presumed daemon-stimulated nucleon decay turned out to be ~$10^{-6}$ s rather than ~$10^{-7}$ s, as this could be expected), our detector is capable of detecting daemons with $V < 30$ km/s [5].

Significantly, the data collected in the DAMA/NaI experiment can also, in their turn, be interpreted in a straightforward (and even quantitative) way within the frame of the daemon paradigm [4,7].

Indeed, the 2-6 keV effective energy range corresponds to the energy of the iodine recoil nuclei knocked out elastically (i.e., with doubling of the velocity) in NaI(Tl) by a supermassive particle moving exactly with $V = 30$-$50$ km/s, while the 1-year period comes from the Earth's crossing of the abovementioned "shadow" produced by SEECHOs. (Moreover, it is worth noting here that the desire to refine the details of detection of iodine recoil nuclei with energies ≤10 keV with a scintillator brought us to the conclusion that ion channeling in NaI(Tl) crystals plays an essential part in determining the efficiency of its light yield [7], a point which had been overlooked in earlier DM experiments; channeling makes the light yield of ions equal to that of electrons of the same energy.) The SEECHO daemon flux calculated from the modulation amplitude $A$, with due account of channeling in the NaI(Tl) crystal, was found to be $f \sim 6 \times 10^{-7}$ cm$^{-2}$s$^{-1}$ [7] for the DAMA/NaI experiment, a figure which is in accord with our earlier estimates [3] and is not at odds with the NEACHO daemon flux measured in our ground-level and underground experiments [6]. It becomes clear now why other experiments did not support the results of DAMA/NaI. In detectors with Si and Ge, the energies of the recoil nuclei knocked out elastically with the doubled velocity are below the sensitivity threshold because of their mass being smaller than that of the iodine nuclei in the DAMA/NaI experiment, while in the LXe experiments channeling does not exist at all.

We believe that on having captured a nucleus (say, Na, S, Zn, I in a scintillator, or Cu, Pb in the DAMA shield), the daemon carries it with itself while staying inside it and disintegrating nucleons one after another [5]; therefore, for a certain time (~$10^{-6}$ s) this complex (daemon/remainder-of-the-captured nucleus, - c-daemon in what follows) should pass through a stage where it will be electrically neutral, i.e., will have zero electric charge. It is apparently during this time that it knocks out a recoil nucleus elastically out of the crystal lattice. The details of the processes involved require naturally further study. It remains, in particular, unclear why the catalysis of these nucleon decays does not produce scintillations [6].

We are going to show below that the recently published DAMA/LIBRA data [8] not only support the conclusions drawn in the DAMA/NaI experiment and their interpretation involving daemons but offer additional interesting information concerning the parameters of daemon interaction with matter.

## 2. An analysis of the part played by the single-hit criterion may decide the issue not in favor of the WIMPs

Application of the so-called single-hit criterion (SHIC) is an essential factor in the DAMA search for WIMPs [1,2,8].

The point is that, by definition, WIMPs have an extremely small cross section of interaction with nucleons (original estimates suggested, e.g., $\sigma \sim 10^{-38}$ cm$^2$ for a Galactic halo WIMP velocity of ~300 km/s [9]; the absence of any positive results in the search lowers this value down to $\sigma \sim 10^{-43}$-$10^{-42}$ cm$^2$). In condensed matter, a WIMP passes without interaction a distance measured in light years. It is assumed therefore that one particle crossing consecutively, say, two scintillators cannot excite scintillations in both of them. This assumption is used efficiently to exclude false events caused, for instance, by cosmic ray muons; one rejects in this way the events accompanied by scintillations occurring simultaneously (i.e., within a certain time interval; in the DAMA/NaI and LIBRA cases it was adopted to be 600 ns [1,8]) in crystals, adjacent or located one under another. This is what the SHIC means.

On the other hand, although daemons pass through the Sun or the Earth (in Baksan in September, for instance, we detect their primary flow from below [10]), this occurs only because due to their giant mass they have originally an immense kinetic energy, which exceeds by far the losses incurred by the resistance of the medium. In particular, as follows from our experiments aimed at their detection, after a capture of the Zn nucleus in the ZnS(Ag) layer a NEACHO daemon regains its activity and becomes capable of exciting a signal sometime after $\Delta t \sim$ 20-40 μs [5,6]. Propagating with a velocity $V$ = 10-15 km/s, it passes during this time in our detector a distance of ~30 cm (see above).

A SEECHO daemon moving with a velocity of ~30-50 km/s should regain its activity after capture of a Zn (or Cu) nucleus on passing a distance of ~1 m, and of a Pb nucleus, a thrice larger path (copper, lead and cadmium-foil are used in the DAMA set-up for its shielding from the outer background radiation [1,2,8]).

One can therefore expect that application of SHIC to a detector of the type of DAMA/NaI containing not 9 (3×3) crystals (vertical size of the system ≈50 cm) but 25 crystals (i.e., 5×5, as in the DAMA/LIBRA, with vertical size of close to 1 m) and, further, 100 (i.e., 10×10 crystals, as in the 1-ton detector planned to be built by the DAMA Collaboration, with a vertical size >1.5 m) will entail a loss of information if these systems detect not WIMPs but other DM objects, say, daemons (this reasoning, which one may consider as a prediction following from the daemon paradigm, can be found in the report submitted in March 2008 at the "Problems of Practical Cosmology (PPC-2008)" Conference; Vol. I of the Conference Proceedings [11] was sent by editors to the printer's on April 14, 2008, i.e., before the opening of the Conference and before the publication of the DAMA/LIBRA results [8]).

## 3. DAMA/LIBRA data likewise favor the existence of daemons

The four-year exposure of DAMA/LIBRA with 24 scintillators has raised the confidence level of existence of the annual signal rate modulation in the 2-6-keV range to 8.2σ, and even to 8.9σ (!), if one limits oneself to the 2-5-keV range [8]. The latter observation, if treated together with the comment that "the difference in the (2-6) keV modulation amplitude between DAMA/NaI and DAMA/LIBRA depends mainly on the rate in the (5-6) keV energy bin" [8] is particularly intriguing.

Indeed, the modulation amplitude $A$ in DAMA/LIBRA observed in the 2-4-, 2-5- and 2-6-keV ranges is, on the general, *systematically lower* than that in DAMA/NaI; in the latter case of 2-6 keV it is nearly one half that of DAMA/NaI and is beyond experimental error ($A$ = 0.0107±0.0019 for DAMA/LIBRA, to be compared with $A$ = 0.0200±0.0032 cpd/kg/keV for DAMA/NaI). This is evident from Fig. 1 in [8], where the distribution of single-hit scintillation events in the 5-6-keV interval does not practically differ from that obtained at energies >6 keV, where no modulation is seen.

A question naturally arises why the 5-6-keV events in DAMA/LIBRA reveal little or no information on annual modulation while events with energies <5 keV still carry it?

The DAMA Collaboration is inclined to believe that this difference is due to purely statistical deviations within the 2σ range [8].

Our answer to this question has actually been given at the end of the preceding Sec. 2, namely, the DAMA detectors detect not WIMPs but rather other DM particles (daemons!?), and, therefore, the SHIC criterion should be dropped. Indeed, a SEECHO daemon, on interacting with matter in one NaI(Tl) crystal and crossing a path of ~0.5-1 m, is capable of knocking out another iodine recoil nucleus in another crystal. Whence it follows, in particular, that the data processing technique accepted in this case simply cannot yield, generally speaking, model-independent conclusions concerning the properties of DM objects.

This prompts immediately, however, another question of why it is close to the upper energy threshold, i.e., in the 5-6-keV bin, that this effect manifests itself so strongly?

In our opinion, the answer to this question is fairly obvious. The neutral c-daemon moving with velocity $V$ knocks out an ion with velocity $2V$ only in a head-on collision. In tangent collisions, which are certainly more probable, the velocity (and energy) of the knocked out ion will be lower.

And if at an iodine recoil energy $E_r \approx 6$ keV imparted in a nearly head-on collision with the c-daemon moving with a velocity $V$ = 50 km/s the next (or preceding) off-central collision in another crystal produces a recoiled ion with $E_r \approx 3$ keV, the SHIC criterion will veto subsequent analysis of this event. But for a daemon with $V \leq 40$ km/s ($E_r \leq 4$ keV) the event will not be dropped, because a similar second (or the first) off-central hit will create a recoil nucleus with $E_r \leq 2$ keV, which is below the sensitivity threshold of the system $E_0$ = 2 keV, - with the result that SHIC will not veto this event.

All the above-said is quite obvious for the case when the time window for the SHIC application is greater than the time of the DM transit through the scintillator assemblage of ~0.5-1 m by dimension. The latter is about 2-3 μs for the Galaxy halo objects ($V \sim 300$ km/s) and ~10-30 μs for the SEECHO objects. That is much greater than the time gate 600 ns exploited by the DAMA experimenters for the SHIC application for proving that these, we say "close", multiple-hits events have no annual modulation which could be due to non-astronomical background causes. However, even at such a narrow time gate a lowering of the high-energy bin event rate must proceed[1]. That depends on what kind of collision occurs the first. If, as in the above $V = 50$ km/s example, the first collision being off-central produces the $E_r \approx 3$ keV signal, in 600 ns it will veto the registration system for 500 μs and the next 6 keV signal occurring some tens μs later will not be recorded. As a consequence, the rate of such registered high-energy bin events per kg of the scintillator mass drops. That does not concern the above mentioned corresponding situation with $V \leq 40$ km/s, - there will be only one signal with $E_r > E_0 = 2$ keV (i.e. with $2 < E_r \leq 4$ keV) and this event will be saved. Generally speaking, the slowly mowing c-daemons find themselves in an advantageous position in a more bulky detector as they have a greater chance to be registered as moving with a velocity close to their real velocity. Such c-daemons are using the greater volume of the detector. This factor also results in a more effective "washing" out of the high-energy (5-6 keV bin in this case) tail of the resulting distribution of events recorded with making use of SHIC. (The above figures are given as an illustration only. The details are defined by a true dependence vs. collision parameter of a cross-section of elastic interaction of the neutral c-daemon with the nucleus being recoiled. This dependence is hardly the same one as for a neutron.)

Thus the registration of the high-velocity part of the SEECHO daemon transits does not use effectively all the scintillators of the enlarged LIBRA system, - in a sense, it is a "surface" process when the SHIC is used. Therefore, in the limit of a very large scintillator assemblage, the annual-modulated rate for 5-6 keV events has to increase proportionally to the exposed surface area (normal to the daemon flux), not to the assemblage mass (one has to remember also that the distribution on velocity of SEECHO objects falling to the Earth is not constant, - its tails at 30 and 50 km/s are not abundant naturally).

So it is obvious that as the bulk size of such a compact detecting system is increased, i.e., if a 10×10 element array is used in the 1-ton NaI(Tl) system under planning, the mean upper detection threshold for $E_r$ and the event rate (measured in cpd/kg) will decrease. It thus follows that in order to detect daemons, one should probably, rather than trying to attain a maximum possible mass×time product (which is justified for WIMPs), develop detectors with a large surface area which would provide a set of large area×time products (an approach that had been pursued in our experiments which used plane thin ZnS(Ag) scintillators; employment of not very thick scintillators makes the SHIC application to be unneeded).

---

[1] A note which could be of interest for psychologists trying to unravel the enigma of creative work. Being completely sure of the validity of the daemon paradigm which is supported by our experiments [5,6], I have been tortured for all of the four or five days and nights by the problem of how to reconcile the practical absence of modulated LIBRA signals in the 5-6-keV bin with the absence of modulation of multiple-hit events, as well as with the narrow time gate of 600 ns for the SHIC used by the DAMA Collaboration. I tried different approaches, one after another. Finally, on the fifth morning, on having waked up but still slumbering, I got it: in 600 ns after the first 2-3-keV signal (a non-head-on collision knocking out of an iodine ion by a particle with $V = 50$ km/s; see text) the system becomes blocked for the subsequent 500 μs and, thus, will not be able to detect a second signal in the 5-6-keV bin which may arrive a few tens of μs later and could possibly be stronger (in the $V \leq 40$ km/s case, this is not so: the system will not respond to a similar non-head-on first signal, because its amplitude is $< E_0 = 2$ keV; see text).

One more intriguing consequence ensues, namely, a lowering of the detector sensitivity threshold to $E_0 < 2$ keV may bring about not an increase but rather a decrease in the number per the energy bin of the 2-6 keV events detected with using SHIC.

## 4. Conclusions

We thus see that the data amassed in the DAMA/NaI and DAMA/LIBRA experiments find a straightforward, and even quantitative interpretation in terms of the daemon paradigm, which apparently cannot be said about the WIMP paradigm.

When combined with our St-Petersburg (and Baksan) experiments, these results suggest the existence in the Solar system of two populations of daemons being captured and accumulating in SEECHOs and NEACHOs. The DAMA/NaI experiment detects a flux $f \sim 6 \times 10^{-7}$ cm$^{-2}$s$^{-1}$ of the first population which falls on the Earth with $V = 30-50$ km/s [4,7]. Our St-Petersburg experiments reveal the presence of a similar flux, $f > 10^{-7}$ cm$^{-2}$s$^{-1}$, of the second population that crosses the Earth with $V = 11.2-15$ km/s [6,10-12].

As pointed out earlier by us [7], correct interpretation of the DAMA experiments hinges upon taking into account the channeling effect of low-energy (≤10 keV) iodine recoil nuclei in NaI(Tl) crystals, which identifies the 2-6-keV signals exactly with the SEECHO velocities of 30-50 km/s. In addition, to match the St-Petersburg with DAMA experiments, one has to assume that the effective cross section of daemon interaction with the solar material is $\sigma \approx 10^{-19}$ cm$^2$ [4]. This is a giant value compared with the figure assumed for the WIMP interaction with nucleons ($\sigma \sim 10^{-43}$-$10^{-42}$ cm$^2$). This immediately suggests a conclusion, which is prompted also by the St-Petersburg experiments, that the SHIC criterion in its simplest version is inapplicable here, and this conclusion is corroborated now by a comparison of the DAMA/NaI with DAMA/LIBRA experiments. This comparison can be used to derive information on the rate and modes of nucleon digestion in the nuclei captured and transported by a daemon (that the modes of daemon-stimulated nucleon decay are not clear followed already from our simple experiments with a massive CsI(Tl) scintillator [6]). In particular, the about twofold decrease in the mean residual amplitude $A$ for LIBRA (if compared with DAMA/NaI) implies the mean path between the effective c-daemon interactions with matter is measured by tens of cm that is consistent with our St-Petersburg experiments [5,6].

Obviously enough, particular attention should be focused on treatment of multiple-hits events, especially of double events with time shifts corresponding to $V = 30-50$ km/s. This will make physically meaningful increasing the sensitivity to cover the region of $E_0 < 2$ keV of electronic equivalent (where one may hope to reveal the SEECHO harmonic with $P = 0.5$ y, and with $E_0 < 0.5$ keV, the NEACHO harmonic with $P = 0.5$ y and $V = 11.2-15$ km/s detected in St. Petersburg experiments). In the discussed DAMA/LIBRA experiment, multiple-hits events have not thus far revealed annual periodicity [8], which can be accounted for, if for no other reason, by their considering in a rather narrow time gate, - 600 ns only [2,13]. Adopting such a small time interval was justified by the only desire to demonstrate that no other side effects (say, from hardware, background, etc) could produce the annual modulation [2]. It should be stressed once more that it is only those binary events whose shift in time between the two scintillations corresponds to 30-50 km/s that should exhibit the periodicity.

The existence of daemons rather than WIMPs provides a simple explanation for why other experiments exploiting different types of detectors do not reproduce the DAMA results. On the other hand, the crossing by NEACHO daemons of the extended gaseous tritium source which is surrounded by superconducting Nb-containing windings in systems aimed at direct measurement of neutrino mass, and emission of Auger electrons by daemon-captured and transported Nb atoms within a narrow energy interval of 18566-18572 eV may account for the so-called Troitsk anomaly, i.e., the drift of the tritium β spectrum tail with a $P = 0.5$ y periodicity [14] (SEECHO daemons cross the source channel cross section too fast to produce a noticeable $P = 1$ y periodicity). This prompts certain predictions concerning observation of the Troitsk anomaly in the future experiment KATRIN [14].

Turning back to the DAMA experiments, one can repeat once more [14] that "attempts at untangling the subtleties of somebody else's experiment is a risky and extremely unrewarding task. Those who are directly involved in this experiment will immediately locate inaccuracies in such an attempt." Nevertheless one can predict with confidence that:

(1) further DAMA/LIBRA experiments will not leave any doubt in the reality of decrease of the annually modulated residual rate for (now) single hit events in the 2-6 keV range and,

(2) of the especially expressive decrease in the 5-6 keV region (at the same $E_0 = 2$ keV software energy threshold),

(3) the double events with the SHIC window of some tens of μs will show the annual modulation,

(4) this, in its turn, will reveal that the DM object detection with employing SHIC in its present form is the "surface" process, and

(5) confirm an idea that the annual modulation of the signal rates observed in DAMA/NaI and DAMA/LIBRA experiments is caused by the SEECHO flux of (not weakly interacting) daemons.

We hope the highly qualified members of the DAMA Collaboration will develop this approach for interpreting the DAMA results as that has occurred earlier [15] with an effect of channeling of low-energy recoil nuclei in NaI(Tl) crystals pointed out by us [7]. Such a theory will permit to reveal new aspects of the daemon interaction with matter.